\documentstyle[11pt]{article}

\title{\bf The EMC effect at low $x$ in perturbative QCD}
\author{N.Armesto and M.A.Braun\thanks{Visiting professor IBERDROLA; on
leave of absence from the  Department of High Energy Physics,
University of St. Petersburg, 198904 St. Petersburg, Russia.}\\ Departamento
de F\'{\i}sica de Part\'{\i}culas,\\ Universidade de Santiago de
Compostela,\\ 15706-Santiago de Compostela, Spain}
\date{March 1996}
\def\beq{\begin{equation}}
\def\eeq{\end{equation}}
\def\noi{\noindent}

\begin{document}
\maketitle
\medskip
\noi{\bf Abstract.}

The EMC effect is studied in the perturbative QCD hard pomeron approach. In
the limit $x\rightarrow 0$ and for a nucleus with a constant density the
effect is found to be a function of a single variable which combines its
$x$-, $Q$- and $A$-dependence. At large $Q^{2}$ the effect dies out as $\ln
Q/Q$. At small but finite $x$ a change in the anomalous dimension leads to a
weaker $Q$-dependence, which almost disappears as $x$ grows and the anomalous
dimension becomes small. In this approach the nuclear structure function can
be expressed directly in terms of the proton one.
Calculations of the EMC effect on these lines give results which agree
reasonably well to the existing experimental data for $x<0.05$.

\vspace{7.5cm}
\noi{\Large\bf hep-ph/9603360}\\
\noi{\Large\bf US-FT/15-96}
\newpage
\section{Introduction}
The nature of the EMC effect at small $x$ is well understood
qualitatively.
At high enough $Q^{2}$ the incoming photon splits into a $q\bar q$ pair
which interacts with the nucleus target in the manner typical for a
hadronic projectile and thus experiences absorption which makes the
structure function (per nucleon) become smaller than for a free nucleon
[1-5].
An alternative, though equivalent, explanation (appropriate for the
reference system in which the nucleus is moving fast) is that the gluonic
clouds created by different nucleons of the nucleus begin to overlap and
the gluons recombine, so that their density becomes smaller [6-8].
However, as far as quantitative
predictions are concerned,  published results  heavily depend on additional
semiphenomenological assumptions and often contradict each other. In
particular the $Q$-dependence of the effect results different in different
approaches. In a serie of papers  [3-5] it is argued that $q\bar q$
configurations of a large dimension give the dominant contribution to the
absorption, which results essentially independent of $Q$.  On the other
hand in the gluon recombination approach of [6-8] the absorption is
obtained as a clear higher-twist effect dying out at large $Q$. 

In this note we calculate the nuclear structure function in
perturbative QCD which seems to be applicable to the low $x$ region. To
simplify the treatment we additionally take the number of colours $N_{c}$
to be very large. In this limit the low $x$ nuclear structure function 
can be calculated in a closed form and results quite simple: it is 
given by the exchange of any number of BFKL pomerons [9] between the
projectile $q\bar q$ pair and nucleons of the target. Corrections have the
relative order of $1/N_{c}^{2}$, which may be not too bad for the physical
case $N_{c}=3$.

A peculiar property of the BFKL pomeron is that, due to the
unperturbative rise of the anomalous dimension up to unity, the effective
cross section of the coloured dipole formed by the $q\bar q$ pair is
proportional to its transverse dimension $r$ rather than to $r^{2}$, as
taken in [3-5]. As a result, a strong violation of scaling is predicted for
the proton structure function $F_{2p}(x,Q^{2})$ at low $x$, which should
rise linearly with $Q$. 

For the nucleus our calculations reveal that  absorptive corrections
due to $n$ interactions with the target ($n\geq 2$) are indeed
independent of $Q$, as stated in [1], and have the order $m^{2-n}$ in the
effective quark mass. However for a heavy nucleus contributions with
different $n$ obtained in this manner cancel in the sum, so that the
resulting absorption results independent of the quark mass. A careful study of
this result shows that contrary to the afore mentioned assertions large
spatial separations of the $q\bar q$ pair are not  singled out, its
typical dimension being small and independent of $Q$, of the order
$A^{-1/3}R_{N}$, where $R_{N}$ is the nucleon radius. For a nucleus with a
constant density inside a sphere of a radius $R_{A}=A^{1/3}R_{0}$ the
calculated EMC effect turns out to be a function of a single variable
\[z_{0}=cA^{1/3}\ \frac{1}{Q}\left(\frac{1}{x}\right)^{\Delta},\]
which combines its $x$-, $Q$- and $A$-dependence. The effect goes down 
as $\ln Q/Q$ at high $Q^{2}$ and rises with $A$ and $1/x$ as expected, its
absolute magnitude remaining relatively small up to $x\sim 10^{-6}$ and not
too low $Q^{2}$.

The asymptotic pomeron approach may only be applicable at quite small
$x<0.001$, at which the existing experimental data on $F_{2p}(x,Q^{2})$
 do not contradict the predicted linear rise with $Q$.
At larger $x$ the growth with $Q$ is less pronounced, which indicates
that the anomalous dimension of the pomeron becomes smaller and tends to
zero at finite $x$. It turns out that the experimental data on
$F_{2p}(x,Q^{2})$  up to
$x<0.05$ can be well described by a subasymptotic pomeron with the
effective anomalous dimension going down logarithmically in $x$. With this
subasymptotic pomeron as an input, we have calculated
the nucleon structure functions for $x<0.05$. Comparison to the existing
experimental data shows  a quite reasonable agreement. 

As a result of a changing anomalous dimension the $Q$ dependence of the
EMC effect results also $x$-dependent, becoming weaker as $x$ grows.
With the anomalous dimension close to zero, the effect becomes
independent of $Q$, in agreement with the statements in [3-5].

\section{The $\gamma^{\ast}A$ scattering in the leading approximation in
$1/N_{c}$}
As is well known, in the high-colour limit all contributions to the
scattering amplitude in perturbative QCD can be classified according
to the topological structure of the corresponding Feynman diagrams [10,11].
The leading diagrams have the topological structure of a cylinder. For
$\gamma^{\ast}$-hadron scattering they sum into a single BFKL pomeron [12].
Diagrams with more complicated topologies correspond to multipomeron
exchanges with all sort of interaction between the pomerons.

For $\gamma^{\ast}$-nucleus scattering the leading diagrams in
$1/N_{c}$ correspond to the exchange of several pomerons with their
possible branchings as they propagate to the nucleus (generalized fan
diagrams, Fig. 1). In the perturbative domain
\beq
g^{2}N_{c}<<1
\eeq
only the simplest of them survive at high energies, those without
branchings (Fig. 1a). Indeed the energy dependence of the two
diagrams of Figs. 1a and 1b is given by $(s_{1}s_{2})^{2\Delta}$ and
$s_{1}^{\Delta}s_{2}^{2\Delta}$ respectively, so that the contribution
from Fig. 1a clearly dominates if $s_{1}$ is large. If, on the other
hand, $s_{1}$ is finite then it continues to be dominant
because in this case the contribution of Fig. 1b includes an extra
factor $g^{2}N_{c}$ not compensated by a large $\ln s$.

Thus in the dominant approximation in $1/N_{c}$ we have to sum only the
diagrams which correspond to Fig. 1a with any number of pomerons. To do
this we have to know the coupling of the projectile photon to an
arbitrary (even) number of gluons, which form colourless pairs. In the
high-colour limit and in the lowest order in $g^{2}N_{c}$ this vertex has
been calculated in [12]. For $2n$ gluons with momenta $q_{i}$,
$i=1,...,2n$, $\sum q_{i}=0$, it is given by 
\beq
F^{(2n)}(q_{i\bot})=ig^{2n}\left(\frac{N_{c}^{2}-1}{2N_{c}}\right)^{n}
\int_{0}^{1}d\alpha\int d^{2}r\ \rho(\alpha,r)
 \prod_{i=1}^{2n}[\exp (irq_{i})-1].
\eeq
\def\aint{\int_{0}^{1}d\alpha\int d^{2}r\ \rho(\alpha,r)}
Here $\rho(\alpha,r)$ is the (dipole) colour density created by the
projectile photon as it splits into a $q\bar q$ pair with a part $\alpha$
of its longitudinal momentum carried by the quark and a transverse
distance $r$ between the quark and the antiquark. For longitudinal ($L$) and
transverse ($T$) photons of virtuality $Q^{2}$ this density has been
calculated in [3]:
\beq
\rho_{L}(\alpha,r)=\frac{4e^{2}N_{c}Q^{2}}{(2\pi)^{3}}
\sum_{f=1}^{N_{f}}Z_{f}^{2}[\alpha
(1-\alpha)]^{2}{\rm K}_{0}^{2}(\epsilon_{f} r),
\eeq
\beq
\rho_{T}(\alpha,r)=\frac{e^{2}N_{c}}{(2\pi)^{3}}
\sum_{f=1}^{N_{f}}Z_{f}^{2}\{m_{f}^{2}{\rm K}_{0}^{2}(\epsilon_{f} r)
+[\alpha^{2}+(1-\alpha)^{2}]\epsilon_{f}^{2}{\rm K}_{1}^{2}
(\epsilon_{f} r)\},
\eeq
where the summation goes over flavours,
\beq
\epsilon_{f}^{2}=\alpha (1-\alpha)Q^{2}+m_{f}^{2}
\eeq
and $m_{f}$ and $Z_{f}$ are the mass and charge of the quark of flavour
$f$. 

The $2n$ gluons coupled to the vertex (2) form $n$ pomerons as a result
of their interaction. We assume that the pomerons are formed by pairs of
gluons ($1,n+1$), ($2,n+2$) and so on. Upon reaching the nucleus each pomeron
is coupled to a nucleon with a vertex analogous to (2) with $n=1$, depending on the
primed momenta $q'_{i}$ and $q'_{n+i}$, $i=1,...,n$, and involving an
unknown (nonperturbative) colour density for the nucleon
$\rho_{N}(\alpha'_{i},r'_{i})$. It is also clear that each pomeron has
its total momentum equal to zero: $q_{i}+q_{n+i}=q'_{i}+q'_{n+i}=0$.
Then its propagator will be given by the BFKL Green function for the
forward scattering [13]:
\beq if(\nu,,q_{i},q'_{i})/(N_{c}^{2}-1),\ i=1,...,n,\eeq
where $\nu=pq,\  q_{n+i}=-q_{i},\ q'_{n+i}=-q'_{i}$. The factor
$1/(N_{c}^{2}-1)$ comes from the projection onto the colourless state.

Integrations over $q_{i}$ and $q'_{i}$ with the exponential factors
contained in the vertices transform this Green function to the coordinate
space. As a function of coordinates, the BFKL Green function vanishes at
the origin, so that terms with unity instead of $\exp (iq_{i}r)$ in (2)
give no contribution. As a result each pomeron propagator turns into the
coordinate Green function for the forward scattering 
\beq 4if(\nu,r,r'_{i})/(N_{c}^{2}-1),\ i=1,...,n, \eeq
which should be integrated over $r$ and all $r'_{i}$ with the
corresponding colour densities.

One has to finally recall that each interaction with a nucleon of the
target is accompanied by a nuclear profile function factor $T(b)$ giving
the probability to meet a nucleon at a given impact parameter $b$. Also a
binomial factor $C_{A}^{n}$ should be included for $n$ interactions.

Then in the end we arrive at the following expression for the 
$\gamma^{\ast}A$ forward scattering amplitude with $n$ interactions:
\beq
A^{(n)}(p,q)=-4i\nu C_{A}^{n}w_{n}\aint [-\sigma(r)/2]^{n}.
\eeq
Here
\beq
\sigma(r)=\frac{2g^{4}}{(2\pi)^{2}}\frac{N_{c}^{2}-1}{N_{c}^{2}}
\int_{0}^{1}d\alpha'\int d^{2}r'\rho_{N}(\alpha',r')
f(\nu,r,r')
\eeq
has the meaning of the cross section for the scattering of a colour
dipole formed by the $q\bar q$ pair of transverse dimension $r$ off
the nucleon. The factor $w_{n}$ gives the total probability to find $n$
nucleons at the same impact parameter:
\beq
w_{n}=\int d^{2}bT^{n}(b).
\eeq

Evidently the amplitude (8) has an eikonal form under the sign of
integration over $\alpha$ and $r$. The  $\gamma^{\ast}A$ total
cross section is obtained  as ${\mbox Im}\,A/(2\nu)$. From (8),
\beq
\sigma_{A}^{(n)}=-2 C_{A}^{n}w_{n}\aint [-\sigma(r)/2]^{n}.
\eeq
Summing over all possible numbers of interactions we find the total
cross section:
\beq
\sigma_{A}=2\int d^{2}b\aint\left\{1-[1-(1/2)\sigma(r)T(b)]^{A}\right\}.
\eeq

Now we turn to the cross section $\sigma(r)$ defined by (9). The
explicit form of the BFKL Green function which enters (9) is [13]
\beq
f(\nu,r,r')=\frac{rr'}{8}\int_{-\infty}^{\infty}d\kappa
\frac{\nu^{\omega (\kappa)}}{(\kappa^{2}+1/4)^{2}}\left (\frac{r'}{r}
\right )^{2i\kappa},
\eeq
where
\beq
\omega(\kappa)=-(g^{2}N_{c}/2\pi^{2})[{\rm Re}\,\psi(1/2+i\kappa)-\psi(1)].
\eeq
We have retained only the isotropic part, which gives the dominant
contribution at high $\nu$. At such $\nu$ small values of $\kappa$ contribute
in (13) for which $\omega(\kappa)$ has the form
\beq
\omega(\kappa)=\Delta-a\kappa^{2},
\eeq
where, in terms of $\alpha_{s}=g^{2}/4\pi$,
\beq
\Delta=(\alpha_{s}N_{c}/\pi)4\ln 2
\eeq
is the pomeron intercept and
\beq
a=(\alpha_{s}N_{c}/\pi)14\zeta(3).
\eeq
With (15) the asymptotical expression for the Green function results
\beq
f(\nu,r,r')_{\nu\rightarrow\infty}
\simeq 2rr'\nu^{\Delta}\sqrt{\frac{\pi}{a\ln\nu}}
\exp\left( -\frac{\ln^{2}(r/r')}{a\ln\nu}\right ).
\eeq

As one may see from (8) and (9) characteristic values of $r$ and $r'$ are of the
order $1/Q$ and $1/m_{N}$ respectively, where $m_{N}$ is the nucleon
mass. Then the exponential factor in (18) is close to unity, unless $Q$
becomes so large that $\ln^{2} Q\sim\ln\nu$, which we assume not to be
the case. Omitting this factor we obtain
\beq
\sigma(r)=\mu r R_{N},
\eeq
where 
\beq
R_{N}=\int_{0}^{1}d\alpha'\int d^{2}r'r'\rho_{N}(\alpha',r')
\eeq is the average transverse dimension of the nucleon and the factor
$\mu$ is given by
\beq
\mu=\frac{16\pi\alpha_{s}^{3/2}}
{\sqrt{14 N_{c}\zeta(3)}}
\frac{N_{c}^{2}-1}{N_{c}^{2}}\frac{\nu^{\Delta}}{\sqrt{\ln\nu}}\ \ .
\eeq
It becomes large as $\nu\rightarrow\infty$.

Eqs. (12) and (19) allow in principle to calculate the $\gamma^{\ast}A$
cross section in a straightforward manner for the longitudinal and
transverse photons using the densities given by (3) and (4). The nuclear
structure function at low $x$ can then be calculated as
\beq
F_{2A}=\frac{Q^{2}}{\pi e^{2}}\left[\sigma_{A}^{(T)}+\sigma_{A}^{(L)}\right].
\eeq
However before doing so, we have to make two important remarks.

The first one concerns the dependence of $\sigma(r)$ on $r$. It is
proportional to $r$ rather than to $r^{2}$ as assumed in 
[3-5]. This fact can be traced to the anomalous behaviour of the
BFKL Green function at small $r$, resulting in its turn from the
nonperturbative rise of the anomalous dimension to unity around the BFKL
singularity point. As a result, a large violation of scaling is
predicted for the structure function at small $x$: it grows linearly with
$Q$. This fact will be clearly seen in what follows.

Our second remarks concerns the validity of the expression (12). It has been
obtained  under the standard assumption that the radius of the strong
interaction is much smaller than that of the nucleus. Technically it
reduces to the strong inequality
\beq
\sigma(r)T(b)<<1.
\eeq
In fact, if $\sigma(r)\sim r_{0}^{2}$, where $r_{0}$ is the effective
interaction radius, then (23) tells us that $r_{0}<<R_{A}$.  With
(19) and (3)-(5) one observes that $\sigma(r)$ can take on large values for
light quarks, of the order $1/m_{f}$, when the standard assumption that the
projectile can interact with only one nucleon at a time becomes wrong.
However, on
physical grounds the expressions (3) and (4) cannot be true for very small
values of $\epsilon_{f}$, which allow for very large values of $r$. At such
distances nonperturbative confinement effects should become notable, which
prevent the $q\bar q$ pair from having too large dimensions, larger than some
maximal possible dimension $r_{max}$ of the order of a typical hadronic
size. Then  Eq. (23) will always be fulfilled. To simulate the confinement
effects we shall substitute the small masses $m_{f}$ of light quarks by an
effective regulator mass $m\sim 1/r_{max}$. 

Finally, to simplify the treatment, for $A>>1$ and using (23) we can,
as usual, transform
(12) into an equivalent expression:
\beq
\sigma_{A}=2\int d^{2}b\aint\left \{1-\exp[-(1/2)A\sigma(r)T(b)]\right\}.
\eeq

\section{The nuclear structure function in the limit $x\rightarrow 0$}
We begin with the main part of the structure function given
by the cross section for the transversal photon. Putting the transversal
photon density (4) and the cross section $\sigma(r)$ given by (19) into
(24) and changing the variable $r\rightarrow r/\epsilon_f$ we
obtain \[ \sigma_{A}^{(T)}=
\frac{2e^{2}N_{c}}{(2\pi)^{3}}
\sum_{f=1}^{N_{f}}Z^{2}_{f}
\int d^{2}b\int_{0}^{1}d\alpha\int d^{2}r\]\beq
\times\left[ \frac{m_{f}^{2}}{\alpha (1-\alpha)Q^{2}+m_{f}^{2}}
{\mbox K}_{0}^{2}(r)+
[\alpha^{2}+(1-\alpha)^{2}]{\rm K}_{1}^{2}
(r)\right ]
\left [1-\exp\left (-\frac{A\mu r R_{N}T(b)}{2
\sqrt{\alpha (1-\alpha)Q^{2}+m_{f}^{2}}}\right)\right].
\eeq

Expression (25) depends on the quark masses $m_{f}$. From its form it is
evident that the dominant contribution to $\sigma_{A}^{(T)}$ comes from the
region of integration in $\alpha$ where
\beq
\alpha (1-\alpha)Q^{2}<(A\mu R_{N}/R_{A}^{2})^{2}
\sim(\nu^{\Delta}A^{1/3}/R_{N})^{2}.
\eeq
In the asymptotic region $x\rightarrow 0$ and/or $A>>1$, to which 
 this and the following sections are devoted, $\nu^{\Delta}A^{1/3}>>1$, so
that the right-hand side of (26) is large. Then large values of $\alpha
(1-\alpha)Q^{2}$ will give the dominant contribution, which correspond to
small transverse dimensions of the $q\bar q$ pair of the order
\beq
r\sim R_{N}/(\nu^{\Delta}A^{1/3}).
\eeq
Evidently the masses $m_{f}$ do not play any role at such small values of
$r$, so that we can safely put $m_{f}=0$ in (25). Then it simplifies to 
\beq
\sigma_{A}^{(T)}=
\frac{2e^{2}N_{c}Z^{2}}{(2\pi)^{3}}
\int d^{2}b\int_{0}^{1}d\alpha\int d^{2}r
[\alpha^{2}+(1-\alpha)^{2}]{\rm K}_{1}^{2}
(r)
\left [1-\exp\left(-\frac{A\mu r R_{N}T(b)}{2Q
\sqrt{\alpha (1-\alpha)}}\right)\right],
\eeq
where
\beq
Z^{2}=\sum_{f=1}^{N_{f}}Z_{f}^{2}
\eeq
is the average quark charge squared.

Note that the cross section on the
free nucleon follows from (28) in the lowest order in $T(b)$ (and divided by
$A$):
\beq
\sigma_{N}^{(T)}=
\frac{e^{2}N_{c}Z^{2}}{(2\pi)^{3}}\frac{\mu  R_{N}}{Q}
\int_{0}^{1}d\alpha\frac{\alpha^{2}+(1-\alpha)^{2}}
{\sqrt{\alpha (1-\alpha)}}
\int d^{2}rr{\rm K}_{1}^{2}(r).
\eeq
Calculating the integrals over $\alpha$ and $r$,
\beq
\sigma_{N}^{(T)}=
\frac{9\pi e^{2}N_{c}Z^{2}\mu  R_{N}}{512 Q}\ \ .
\eeq
It is evident that at high $Q$ the cross section $\sigma_{A}^{(T)}\simeq
A\sigma_{N}^{(T)}$ so that the EMC effect dies out.

It is instructive to compare the total cross section (28) with separate
contributions coming from  a given number of interactions $n$. From (11) we
get:  \[
\sigma_{A}^{(T,n)}=
-\frac{2e^{2}N_{c}}{(2\pi)^{3}}C_{A}^{n}w_{n}\sum_{f=1}^{N_{f}}Z^{2}_{f}
\int_{0}^{1}d\alpha\int d^{2}r\]\beq
\times\left[ \frac{m_{f}^{2}}{\alpha (1-\alpha)Q^{2}+m_{f}^{2}}
{\mbox K}_{0}^{2}(r)+
[\alpha^{2}+(1-\alpha)^{2}]{\rm K}_{1}^{2}
( r)\right ]
\left (-\frac{\mu r R_{N}}{2
\sqrt{\alpha (1-\alpha)Q^{2}+m_{f}^{2}}}\right)^{n}.
\eeq
One observes that all terms except the first one diverge as
$m_{f}\rightarrow 0$. The $n=1$ term is independent of $m_{f}$ in this
limit and behaves like $1/Q$ (see (31)). Other terms have the order
\beq
\sigma_{A}^{(T,2)}\sim Q^{-2}\ln (Q^{2}/m_{f}^{2}),\ 
\sigma_{A}^{(T,n>2)}\sim Q^{-2}m_{f}^{2-n}.
\eeq
Thus they are all of the same order $Q^{-2}$ with respect to $Q$ and
therefore asymptotically smaller than the dominant term with $n=1$. For
the structure function (22) they give contributions essentially independent
of $Q$. In this respect the
conclusions in [5] are correct. However comparison to the total
cross section (28), which is independent of $m_{f}$, demonstrates that the
rescattering terms taken separately have little relation to the physical
cross section. Evidently all their terms leading in $1/m_{f}$ cancel in
the sum. The final expression (28) does not admit developing in the
multiple scattering series.

To simplify the calculation of the cross section (28) we choose a
simplified nuclear profile function $T(b)$ corresponding to a finite
nucleus with a constant density
\beq
T(b)=(2/V_{A})\sqrt{R_{A}^{2}-b^{2}},
\eeq
where $R_{A}=A^{1/3}R_{0}$ and $V_{A}$ is the nuclear volume. Then the
integration over $b$ can be done explicitly  and we obtain
\beq
\sigma_{A}^{(T)}=
(2e^{2}N_{c}Z^{2}R_{A}^{2}/\pi)\ J_{T},
\eeq
where $J_{T}$ is given by an integral over $\alpha$ and $r$:
\beq
 J_{T}=\int_{0}^{1}d\alpha\int_{0}^{\infty} rdr
\alpha^{2}{\rm K}_{1}^{2}(r)\chi(z).
\eeq
Here
\beq
\chi(z)=\frac{1}{2}-\frac{1}{z^{2}}+e^{-z}\left(\frac{1}{z}+
\frac{1}{z^{2}}\right)
\eeq
and
\beq
z=\frac{3}{4\pi}\frac{A^{1/3}\mu r R_{N}}{QR_{0}^{2} \sqrt{\alpha
(1-\alpha)}} \equiv
z_{0}\frac{r}{\sqrt{\alpha (1-\alpha)}}\ \ .
\eeq

The cross section for the longitudinal photon can be studied in a similar
manner with the colour density given by (3). In the limit
$m_{f}\rightarrow 0$ it is given by a formula analogous to (28): 
\beq
\sigma_{A}^{(L)}=
\frac{8e^{2}N_{c}Z^{2}}{(2\pi)^{3}}
\int d^{2}b\int_{0}^{1}d\alpha\int d^{2}r
\alpha (1-\alpha){\rm K}_{0}^{2}
( r)
\left [1-\exp\left(-\frac{A\mu r R_{N}T(b)}{2Q
\sqrt{\alpha (1-\alpha)}}\right)\right],
\eeq
and possesses the same
properties and behavior in $Q$ as $\sigma_{A}^{(T)}$. For a free nucleon
the longitudinal cross section turns out to be
\beq
\sigma_{N}^{(L)}=
\frac{\pi e^{2}N_{c}Z^{2}\mu  R_{N}}{256 Q}\ \ .
\eeq
Finally, with the nuclear profile function chosen according to (34) we
obtain
\beq
\sigma_{A}^{(L)}=
(4e^{2}N_{c}Z^{2}R_{A}^{2}/\pi)\ J_{L},
\eeq
where now
\beq
 J_{L}=\int_{0}^{1}d\alpha\int_{0}^{\infty} rdr
\alpha (1-\alpha){\rm K}_{0}^{2}(r)\chi(z),
\eeq
with $\chi(z)$ and $z$  defined by (37) and (38).

According to (35)-(38) and (41),(42) the EMC effect turns out to depend on
a single variable $z_{0}$ which combines its $x$-, $Q$- and $A$-dependence.
Indeed  for the ratio $R^{(A)}(x,Q^{2})$ of the
nuclear structure function per nucleon to the proton one we find
\beq
R^{(A)}(x,Q^{2})\equiv\frac{F_{2A}(x,Q^{2})}{AF_{2N}(x,Q^{2})}=
R(z_{0}),
\eeq
where, according to (38),
\beq
z_{0}=\frac{3}{4\pi}\frac{A^{1/3}\mu R_{N}}{QR_{0}^{2}}\ \ .
\eeq
In principle, $z_{0}$ may take on arbitrary values, from very
small ones up to very large ones. With the growth of $z_{0}$ the effect (i.e.,
$1-R^{(A)}$) also grows  from zero to unity (Fig. 2). At small $z_{0}$ it
behaves like $z_{0}\ln z_{0}$ (see Appendix), which corresponds to going
down like $\ln Q/Q$ at high $Q^{2}$.

To find physically relevant values of $z_{0}$ we note that  
using (22), (31) and (40) we find the proton structure function:
\beq
F_{2p}(x,Q^{2})=(11/512)N_{c}Z^{2}\mu R_{N}Q.
\eeq
As mentioned in the Introduction, it exhibits a strong violation of scaling:
it rises linearly with $Q$, which means that the anomalous dimension has
risen from its presumably small value up to unity in the asymptotical
region. Eq. (45) may be used to express the unknown product $\mu R_{N}$ in 
(44) via $F_{2p}(x,Q^{2})$ to relate the parameter $z_{0}$ to the observable
proton structure function:
\beq
z_{0}=\frac{384}{11\pi}\frac{A^{1/3}}{N_{c}Z^{2}Q^{2}R_{0}^{2}}
F_{2p}(x,Q^{2}).
\eeq
This allows to calculate the nuclear structure function entirely in terms
of the proton one without any unknown parameter,
\beq
R^{(A)}(x,Q^{2})=
\frac{2N_{c}Z^{2}R_{A}^{2}Q^{2}}{\pi^{2}AF_{2p}(x,Q^{2})}(J_{T}+2J_{L}).
\eeq
Here the integrals $J_{T,L}$ are calculated
according to (36) and (42) with $z_{0}$ given by (46).

To apply these asymptotic formulas to realistic nuclear structure
functions at given $x$ and $Q^{2}$ we have first to check that the proton
structure function is reasonably well described by the asymptotic 
expression (45). 
If we take $\nu\sim 1/x$ in $\mu$ (Eq. (21)) then it leads to \beq
F_{2p}(x,Q^{2})=c\frac{Qx^{-\Delta}}{\sqrt{\ln(1/x)}}\ \ ,
\eeq
with
\beq
c=\frac{11\pi}{32}\sqrt{\frac{N_{c}}{14\zeta(3)}}
\frac{N_{c}^{2}-1}{N_{c}^{2}}\alpha_{s}^{3/2}
Z^{2}R_{N}.
\eeq
Comparing (48) with the experimental behaviour of $F_{2p}(x,Q^{2})$
one may hopefully determine $\Delta$ and $c$ and from (16) and (49) 
find the parameters $\alpha_{s}$ and $R_{N}$. (Note, however, that
 the exact scale factor which divides $\nu$ in expressions
like $\nu^{\Delta}$ or $\ln\nu$  cannot be determined in the lowest order
BFKL theory. Therefore the linear dependence on $Q$ in (48) may be
modified by terms of the order $\alpha_{s}\ln Q$, which
are assumed small in the theory, but are not so small for realistic
values of $\alpha_{s}$.)

Studying the existing experimental data for $F_{2p}(x,Q^{2})$ at low $x$
[14-17], we find that the simple  formula (48) describes them quite well for
$x<0.001$ (Fig. 3),
with the parameters $\Delta$ and $c$ chosen according to
\beq
\Delta=0.377,\ \ c=0.0536.
\eeq
They correspond to $\alpha_{s}=0.14$ and $R_{N}=0.44$
fm. Thus the asymptotic hard pomeron, at least, does not contradict the
experimental data for the proton structure function for $x<0.001$ and leads
to physically reasonable values of $\alpha_{s}$ and $R_{N}$. This provides a
justification to apply it to the nuclear structure function in the same
region.

Taking (48) for $F_{2p}$ and using (46) and (47) with four flavours and
$R_{0}=1.35$ fm we have
calculated the ratios $R^{(A)}$ for $10^{-6}<x<10^{-3}$, $Q^{2}=2.5\div
100$ GeV$^{2}$
 and $A=12\div208$. Our results are shown by solid curves in Figs. 4-6, which
illustrate the $x$-, $Q$- and $A$-dependence of the EMC effect
respectively. The effect rises with $1/x$ due to the rise of $F_{2p}$, falls
with $Q$ as $\ln Q/Q$  and, naturally, rises with $A$ as $A^{1/3}$.
Its overall magnitude remains not very large in the whole region explored
($1-R^{(A)}<0.5$).

Unfortunately there are no experimental data on the nuclear structure
functions at such low $x$.  To be able to compare our results with the
existing data, we thus have to move to larger $x$, where the asymptotic
formula (48) does not work.

\section{Realistic $x$ and $A$}

In this section we want to consider the region of not so small $x$ and not
so large $A$, where neither the asymptotic form (48) for the proton structure
function is valid nor the right-hand side of (26) is large.

The latter condition can easily be taken into account retaining
in $\epsilon^{2}_f$, Eq. (5), the quark masses squared $m_{f}^{2}$ (with those
for the light quarks substituted by the regulator mass squared $m^{2}$).
Now both terms in $\epsilon_{f}$ (Eq. (5)) become comparable. It means, of
course, that now the dominant configurations of the $q\bar q$ pair may reach
a size of the order $1/m$, i.e., a typically hadronic size.

As to the proton structure function, we have to take into account that its
rise with $Q$ becomes considerably slower at not so small $x$, which
implies that the effective anomalous dimension of the pomeron is, in fact,
$x$-dependent and goes down with the growth of $x$. Of course one may try to
study the behaviour of the pomeron at lower energies (higher $x$) using the
BFKL equation with some appropriately chosen boundary conditions (or a
driving term). However, since these are poorly known, the resulting
predictions are not reliable. Besides, at not too small $x$ subdominant
terms in $1/\ln x$ may become essential, which are beyond any control (see
however [18], where  such terms  are introduced from some theoretical
considerations). For these reasons, rather than try to determine the
subasymptotical behaviour of the pomeron on (poorly known) theoretical
grounds, we use a simple extension of the asymptotic formula (48) to finite
$x$ which only takes into account a change in the anomalous dimension, to be
determined from the comparison to the experimental data. Namely, we assume
that the effective anomalous dimension of the pomeron is $x$ dependent and
smaller than unity at finite $x$. Correspondingly we take instead of (19)
\beq
\sigma(r)=\xi(\nu)(m_{N}r)^{\beta(\nu)}/m_{N}^{2},
\eeq
where $\beta(\infty)=1$ and $\beta(\nu)>1$ at finite $\nu$
(perturbatively $\beta\simeq 2$). We have also introduced the nucleon
mass $m_{N}$ in (51) as a natural scale. Evidently as
$\nu\rightarrow\infty$ (51) goes into (19) with
\beq
\xi(\nu)=\mu(\nu) m_{N}R_{N}.
\eeq

With the cross section $\sigma(r)$ chosen according to (51)
and the quark masses  $m_{f}$ retained, our formulas 
for the nuclear cross sections (35)-(38), (41) and (42) slightly
change because of the flavour dependence. Instead of (35) and (41) we now
find
\beq
\sigma_{A}^{(T,L)}=
(2^{n}e^{2}N_{c}R_{A}^{2}/\pi)\sum_{f=1}^{N_{f}}Z^{2}_{f}J_{T,L}^{(f)},
\eeq
where $n=1,2$ for $T,L$ respectively and the integrals $J_{T,L}$, depending
on $f$, are now calculated with
\beq
z=\frac{\xi (m_{N}r)^{\beta}AR_{A}}
{m_{N}^{2}V_{A}Q^{\beta}[\alpha(1-\alpha)+m_{f}^{2}/Q^{2}]^{\beta/2}}\equiv
z_{0}\frac{r^{\beta}}
{[\alpha(1-\alpha)+m_{f}^{2}/Q^{2}]^{\beta/2}}
\eeq
and the factor $\alpha(1-\alpha)$ in (42)  substituted
as \beq
\alpha(1-\alpha)\longrightarrow\frac{[\alpha(1-\alpha)]^{2}}
{\alpha(1-\alpha)+m_{f}^{2}/Q^{2}}\ \ .
\eeq
The final EMC ratio $R^{(A)}$ is  now calculated according to
\beq
R^{(A)}(x,Q^{2})=
\frac{2N_{c}R_{A}^{2}Q^{2}}{\pi^{2}AF_{2p}(x,Q^{2})}
\sum_{f=1}^{N_{f}}Z_{f}^{2}\left[J_{T}^{(f)}+2J_{L}^{(f)}\right].
\eeq
Evidently now it depends on two variables: $z_{0}$ defined by
(54) and $Q^{2}$.

Calculating the proton structure function with (51) we find instead of (45)
\beq
F_{2p}(x,Q^{2})=(11/512)\lambda N_{c}\xi (Q/m_{N})^{2-\beta},
\eeq
where $\lambda$ is a  factor
resulting from the $\alpha$ and $r$ integrations (and normalized to unity for
$\beta=1$ and $m_{f}=0$).
To find it one has to calculate the integrals $J_{T,L}$ with the function
$\chi (z)$ substituted by its lowest order term $(1/3)z$. If we denote the
results ${\tilde J}_{T,L}^{(f)}$, then
\beq
\lambda =\frac{768}{11\pi^{3}z_{0}}
\sum_{f=1}^{N_{f}}Z_{f}^{2}\left[{\tilde J}_{T}^{(f)}+2{\tilde J}_{L}^{(f)}
\right].
\eeq
It is evidently independent of $z_{0}$ and can be calculated, say, with
$z_{0}=1$.

If we retain the form of $\xi(\nu)$ which
follows from (52) then (57) turns into
\beq
F_{2p}(x,Q^{2})=c\frac{Q^{2-\beta(x)}x^{-\Delta}}{\sqrt{\ln(1/x)}}\ \ ,
\eeq
with $c$ a slowly varying function of $\ln x$, which we take a constant.
Choosing at small but finite $x$ the anomalous dimension in the form
\beq
\gamma(x)=2-\beta(x)=1+a/\ln x+ b/\ln^{2}x,
\eeq
we obtain a good agreement with all the experimental data on the
proton structure function for $x<0.05$ [14-17] with the parameters 
\beq
\Delta=0.237,\ \ c=0.250,\ \ a=3.87,\ \ b=4.62
\eeq
(see Figs. 7,8).
The formulas (59) and (60) then reasonably well extrapolate the asymptotic
pomeron to larger values of $x$, the power $\beta(x)$ taking into account
the decrease of the anomalous dimension.

Comparing (54) and (57) we observe that the nuclear structure function can
again be written exclusively in terms of the proton one. Indeed the
variable $z_{0}$ in (54) can be expressed similarly to (46)
\beq
z_{0}=\frac{384}{11\pi\lambda }\frac{A^{1/3}}
{N_{c}Q^{2}R_{0}^{2}}F_{2p}(x,Q^{2}),
\eeq
with an extra factor $\lambda$ in the denominator defined according to (58).

Using (62) and $\beta$ given by (60) and (61), we have calculated the
EMC ratio $R^{(A)}$ for values of $x<0.05$, $Q^{2}>1$
GeV$^{2}$ and $A>>1$, at which the experimental data are available [19-21].
Four quark flavours have been taken into account (the inclusion of the
$c$ quark results important). The
value of the internucleon distance has been taken $R_{0}=1.35$ fm.
The best agreement with the data is achieved
for  values of the regulator mass $m$ for light quarks of the order of
$0.5\div 1.5$ GeV. The results depend very little on $m$ inside this
interval: the overall change in $1-R^{(A)}$ does not exceed 10\%. The
results for $m=0.8$ GeV together with the data are presented in the Table. 
As one observes, a very reasonable agreement is achieved, taking into
account that the data, as a rule, are averaged over considerable intervals
of $x$ and $Q^{2}$. 

At smaller $x$ the behaviour of $R^{(A)}$ with
$F_{2p}(x,Q^{2})$ given by (59) naturally repeats the one obtained with
(48) in the preceding section, except at low $Q^{2}$ where the effect of 
(effective) quark masses is felt.
This is illustrated in Figs. 4-6, where the $x$-,$Q$- and $A$-dependence of
$R^{(A)}$  calculated with $F_{2p}$ given by (59) is shown by dashed
curves, to be compared with the one obtained with (48) (solid curves).

\section{Conclusions}

As is well known there are serious problems of principle associated with
the BFKL hard pomeron, related to the unitarity and the running coupling.
In spite of all this, the hard pomeron approach remains the only one, based on
first principles, which allows to study low $x$ phenomena, although in a
model theory.

As we have found, the hard pomeron, at least, does not contradict the
superlow $x$ data for the proton structure function ($x<0.001$). With some
modification the pomeron description can be extended to the region of
$x<0.05$. This gives a motivation to apply it to the nuclear structure
function at these $x$.

With an additional assumption of small interaction between pomerons, which
can be justified in the high $N_{c}$ limit, a closed formula is obtained
for the nuclear structure function, expressing it directly in terms of the
proton one. For a pure BFKL pomeron and a nucleus of constant density the
EMC effect results a function of a single variable $z_{0}$, which combines
its $x$-, $Q$- and $A$-dependence. In principle $z_{0}$ can take on arbitrary
values, so that $R^{(A)}$ can vary from unity to zero. However for realistic
$x$, $Q$ and $A$ the variable $z_{0}$ is not so large, so that $R^{(A)}$
remains not very different from  unity, except for
the lowest $x$ and $Q^{2}$ and highest $A$.

A novel feature of the hard pomeron approach is a changing anomalous
dimension. As a result, the behaviour of the EMC effect on $Q^{2}$ depends
on the value of $x$. For very small values of $x$ the effect dies out as
$\ln Q/Q$. However for larger $x$ the $Q$-dependence becomes less
pronounced until it practically disappears as the anomalous dimension
becomes small.

Comparison to the available experimental data for $x<0.05$ shows a
reasonable agreement. The agreement is not ideal, which may be explained
partly by the imperfection of the existing data, averaged over
considerable intervals of $x$ and $Q$, partly by the presence of other
subasymptotic terms different from the pomeron, which should be notable
for not very small $x$. Evidently more data at lower $x$ are needed to
finally conclude about the relevance of the hard pomeron for the EMC
effect. 

\section{Acknowledgements}
The authors express their deep gratitude to Prof. Carlos Pajares for his
constant interest in the present work and helpful discussions. N.A. and
M.A.B.thank the CICYT of Spain and
IBERDROLA, respectively,   for financial support.

\section{Appendix: The asymptotics of the EMC effect at high $Q^{2}$}

With the quark mass $m_{f}$ retained and at high $Q^{2}$ the EMC ratio
$R^{(A)}$ is a function of two small variables,
$R^{(A)}=R^{(A)}(z_{0},1/Q^{2})$. Since $R^{(A)}(0,1/Q^{2})=1$
identically, the leading correction term is provided by $R^{(A)}(z_{0},0)$,
that is, we can safely put $m_{f}=0$ to study it.

The regions of small $\alpha$ and $1-\alpha$ are responsible for the leading
behaviour at $z_{0}\rightarrow 0$. Therefore we can limit ourselves with the
transversal cross section and consequently with the integral $J_{T}$, Eq.
(36). Transforming to the integration variable $w$ by
\beq
\alpha (1-\alpha)=1/(w^{2}+4),
\eeq
the dominant contribution will come from large $w$. Choosing $w_{0}>>2$, we
find that at small $z_{0}$
\beq
J_{T}\sim 2\int_{0}^{\infty}rdr{\mbox K}_{1}^{2}(r)
\int_{w_{0}}^{\infty}\frac{dw}{w^{3}}\chi (z),\eeq
where
\beq
z=z_{0}(wr)^{\beta},\ \ 1\leq\beta<2.
\eeq
Passing  to the integration over $z$, we present the right-hand side as
\beq
J_{T}\sim (2/\beta)z_{0}^{2/\beta}\int_{0}^{\infty}r^{3}dr
{\mbox K}_{1}^{2}(r) \int_{z_{1}}^{\infty}\frac{dz}{z^{1+2/\beta}}\ \chi
(z),\eeq
with $z_{1}=z_{0}(w_{0}r)^{\beta}$.

The internal integral over $z$ can be transformed into
\beq
\int_{z_{1}}^{\infty}\frac{dz}{z^{1+2/\beta}}\chi
(z)=\frac{z_{1}^{-2/\beta}} {2+2/\beta}\left[\chi
(z_{1})+(\beta/2)z_{1}F(z_{1})\right], \eeq
where 
\beq
F(z)=\int_{1}^{\infty}dxx^{-2/\beta}\exp (-zx)
\eeq
can be expressed via the incomplete gamma function. Evidently the first
term in (67) gives a contribution to $J_{T}$ regular st $z_{0}=0$ and so a
correction to $R^{(A)}$ of the order $z_{0}\sim 1/Q$. The leading correction
comes from the term with $F(z_{1})$. Straightforward estimates give that at
small $z$
\[
F(z)\sim a+bz^{2/\beta-1},\ \ 1<\beta<2;
\]
\beq
F(z)\sim a+bz\ln z,\ \ \beta=1.
\eeq

Putting this into (67) and then into (66) we obtain our final result. 
At high
$Q^{2}$ \[R^{(A)}\simeq 1-c\ Q^{1-2/\beta},\ \ 1<\beta<2;\]
\beq
R^{(A)}\simeq 1-c\ln Q/Q,\ \ \beta=1.
\eeq
In particular, with a  dimension close to its canonical value $\beta=2$ the
EMC effect results independent of $Q$.

\newpage
\section{References}
\noi 1. S.Brodsky, T.E.Close and J.F.Gunion, Phys. Rev. {\bf D6} (1972)
177.\\ 
2. S.Brodsky and H.J.Liu, Phys. Rev. Lett. {\bf 64} (1990) 1342.\\
3. N.N.Nikolaev and B.G.Zakharov, Z. Phys. {\bf C49} (1991) 607.\\
4. V.Barone, M.Genovese, N.N.Nikolaev, E.Predazzi and B.G.Zakharov,
Z. Phys. {\bf C58} (1993) 541.\\
5. B.Z.Kopeliovich and B.Povh, Phys. Lett. {\bf B367} (1996) 329.\\
6. A.H.Mueller and J.Qiu, Nucl. Phys. {\bf B268} (1986) 427.\\
7. J.Qiu, Nucl. Phys. {\bf B291} (1987) 746.\\
8. E.L.Berger and J.Qiu, Phys. Lett. {\bf B206} (1988) 42.\\
9.  V.S.Fadin, E.A.Kuraev and L.N.Lipatov, Phys. Lett. {\bf B60} (1975)
50; I.I.Balitsky and L.N.Lipatov, Sov. J. Nucl. Phys. {\bf 15} (1978) 438.\\
10. G.t'Hooft, Nucl. Phys. {\bf B72} (1974) 461.\\
11. G.Veneziano, Phys. Lett. {\bf B52} (1974) 220; Nucl. Phys. {\bf
B117} (1976) 519.\\
12. M.A.Braun, Univ. of St. Petersburg preprint SPbU-IP-1995/3
(hep-ph/9502403) (to be published in Z. Phys. {\bf C}).\\
13. L.N.Lipatov, in {\it Perturbative Quantum Chromodynamics},
Ed. A.H.Mueller, Advanced Series
on Directions in High Energy Physics, World Scientific, Singapore 1989.\\
14. NM Collab., P.Amaudruz et al., Phys. Lett. {\bf B295} (1992) 159.\\
15. ZEUS Collab., M.Derrick et al., Z. Phys. {\bf C69} (1996) 607.\\
16. ZEUS Collab., M.Derrick et al., Z. Phys. {\bf C65} (1995) 379.\\
17. H1 Collab., T.Ahmed et al., Nucl. Phys. {\bf B439} (1995) 471.\\
18. J.Kwieci\'nski, A.D.Martin and P.J.Sutton, Univ. of Durham preprint
DTP/96/02 (hep-ph/9602320).\\
19. E665 Collab., M.R.Adams et al., Phys. Rev. Lett. {\bf 22} (1992) 3266.\\
20. EM Collab., J.Ashman et al., Z. Phys. {\bf C57} (1993) 211.\\
21. NM Collab. M.Arneodo et al., Nucl. Phys. {\bf B441} (1995) 5; 12.\\

\newpage
\begin{center}
{\Large{\bf Table}}\\
\vspace{0.5 cm}
EMC ratios at experimentally known points.

\vspace{1.5 cm}

\begin {tabular}{|r|r|r|r|r|r|}\hline
   $A$ & $x$ & $Q^{2}$ GeV$^{2}$&   $R$   & $R_{exp}$ & Ref\\\hline
      131 & 0.0065&  1.34 &  0.780 &  0.840$\pm$ 0.107&[19] \\\hline
      131 & 0.0095&  1.91 &  0.825 &  0.800$\pm$ 0.108&[19] \\\hline
      131 & 0.0210&  3.58 &  0.889 &  0.900$\pm$ 0.110&[19] \\\hline
       64 & 0.0310&  4.50 &  0.925 &  0.930$\pm$ 0.033&[20] \\\hline
       64 & 0.0500&  8.50 &  0.955 &  0.955$\pm$ 0.028&[20] \\\hline
       64 & 0.0150&  4.50 &  0.923 &  0.857$\pm$ 0.032&[20] \\\hline
       64 & 0.0310&  3.30 &  0.904 &  0.963$\pm$ 0.022&[20] \\\hline
       64 & 0.0500&  6.40 &  0.944 &  1.005$\pm$ 0.018&[20] \\\hline
       12 & 0.0055&  1.10 &  0.861 &  0.904$\pm$ 0.013&[21] \\\hline
       12 & 0.0085&  1.60 &  0.894 &  0.939$\pm$ 0.012&[21] \\\hline
       12 & 0.0125&  2.20 &  0.917 &  0.939$\pm$ 0.009&[21] \\\hline
       12 & 0.0175&  2.90 &  0.934 &  0.957$\pm$ 0.010&[21] \\\hline
       12 & 0.0250&  3.60 &  0.945 &  0.963$\pm$ 0.009&[21] \\\hline
       12 & 0.0350&  4.50 &  0.954 &  0.990$\pm$ 0.009&[21] \\\hline
       12 & 0.0450&  5.50 &  0.962 &  0.983$\pm$ 0.010&[21] \\\hline
       40 & 0.0085&  1.40 &  0.840 &  0.846$\pm$ 0.014&[21] \\\hline
       40 & 0.0125&  1.90 &  0.871 &  0.870$\pm$ 0.011&[21] \\\hline
       40 & 0.0175&  2.50 &  0.895 &  0.908$\pm$ 0.011&[21] \\\hline
       40 & 0.0250&  3.40 &  0.917 &  0.946$\pm$ 0.009&[21] \\\hline
       40 & 0.0350&  4.70 &  0.937 &  0.956$\pm$ 0.009&[21] \\\hline
       40 & 0.0450&  5.70 &  0.946 &  0.986$\pm$ 0.011&[21] \\\hline
\end{tabular}
\end{center}

\newpage

\begin{center}
{\Large\bf Figure captions}
\end{center}

\noi {\bf Fig. 1.} 
Pomeron diagrams for the EMC effect in the leading approximation
in $1/N_{c}$. With a small coupling constant the diagram (a) dominates at large
energies.

\noi {\bf Fig. 2.} EMC effect as a function of the parameter $z_{0}$ (Eq. (44)).

\noi {\bf Fig. 3.} 
The pomeron fit (48),(50) to the proton structure function for
$x<0.001$ compared to the experimental data [14-17] at $Q^{2}=$ 3, 12 and 35
GeV$^2$.

\noi {\bf Fig. 4.} The $x$-dependence of the EMC effect calculated with the 
proton structure function described by the pure pomeron (Eq. (48), solid
curves) and the subasymptotic pomeron (Eq. (59), dashed curves) for
$x<0.001$, $A=64$ and $Q^{2}=3$ and  50 GeV$^{2}$.

\noi {\bf Fig. 5.} The $Q^{2}$-dependence of the EMC effect calculated with the 
proton structure function described by the pure pomeron (Eq. (48), solid
curves) and the subasymptotic pomeron (Eq. (59), dashed curves) for
$A=64$ and $x=10^{-3}$ and $10^{-6}$.

\noi {\bf Fig. 6.} The $A$-dependence of the EMC effect calculated with the 
proton structure function described by the pure pomeron (Eq. (48), solid
curves) and the subasymptotic pomeron (Eq. (59), dashed curves) for
$x=10^{-3}$ and $10^{-6}$ and $Q^{2}=10$ GeV$^{2}$.

\noi {\bf Fig. 7.} 
The subasymptotic pomeron fit (59)-(61) to the proton structure
function for $x<0.05$ compared to the experimental data [14-17] at
$Q^{2}=3$, 12 and 35 GeV$^2$.

\noi {\bf Fig. 8.} The effective anomalous dimension $\gamma=2-\beta$ as a
function of $x$ (Eqs. (60),(61)).

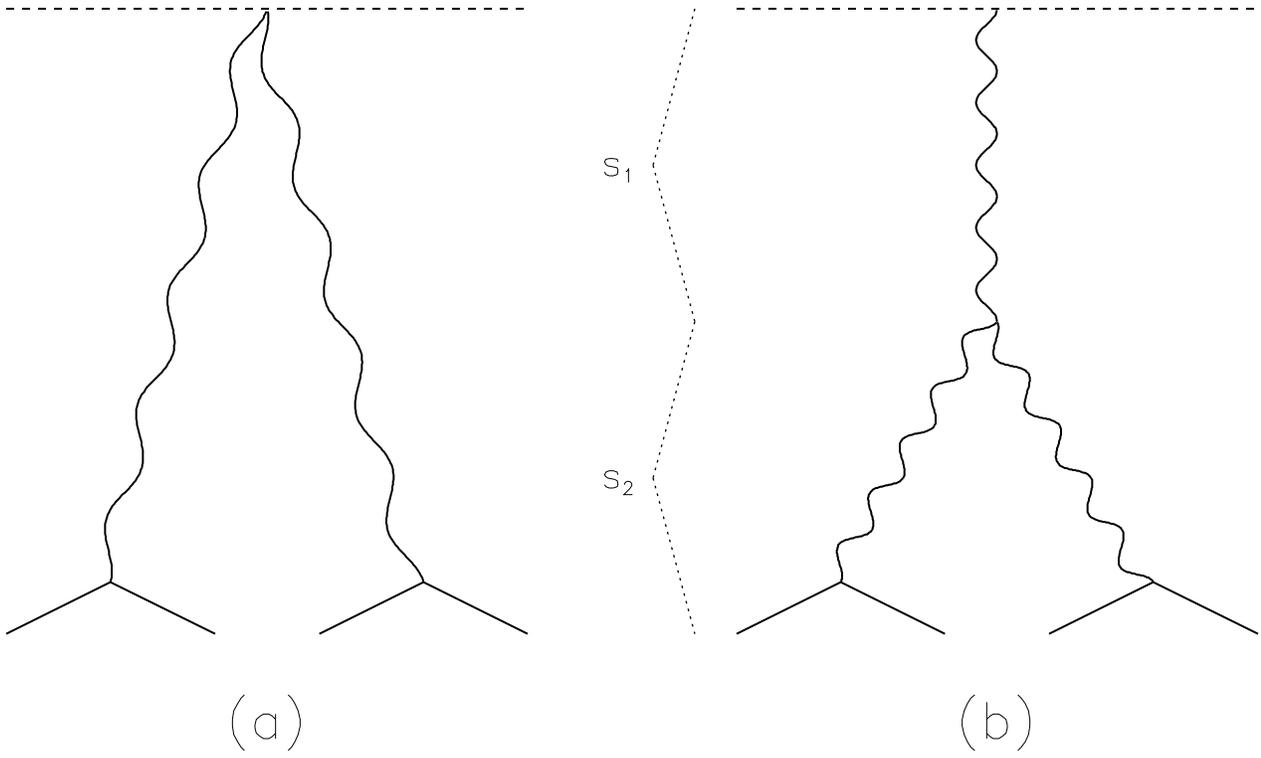
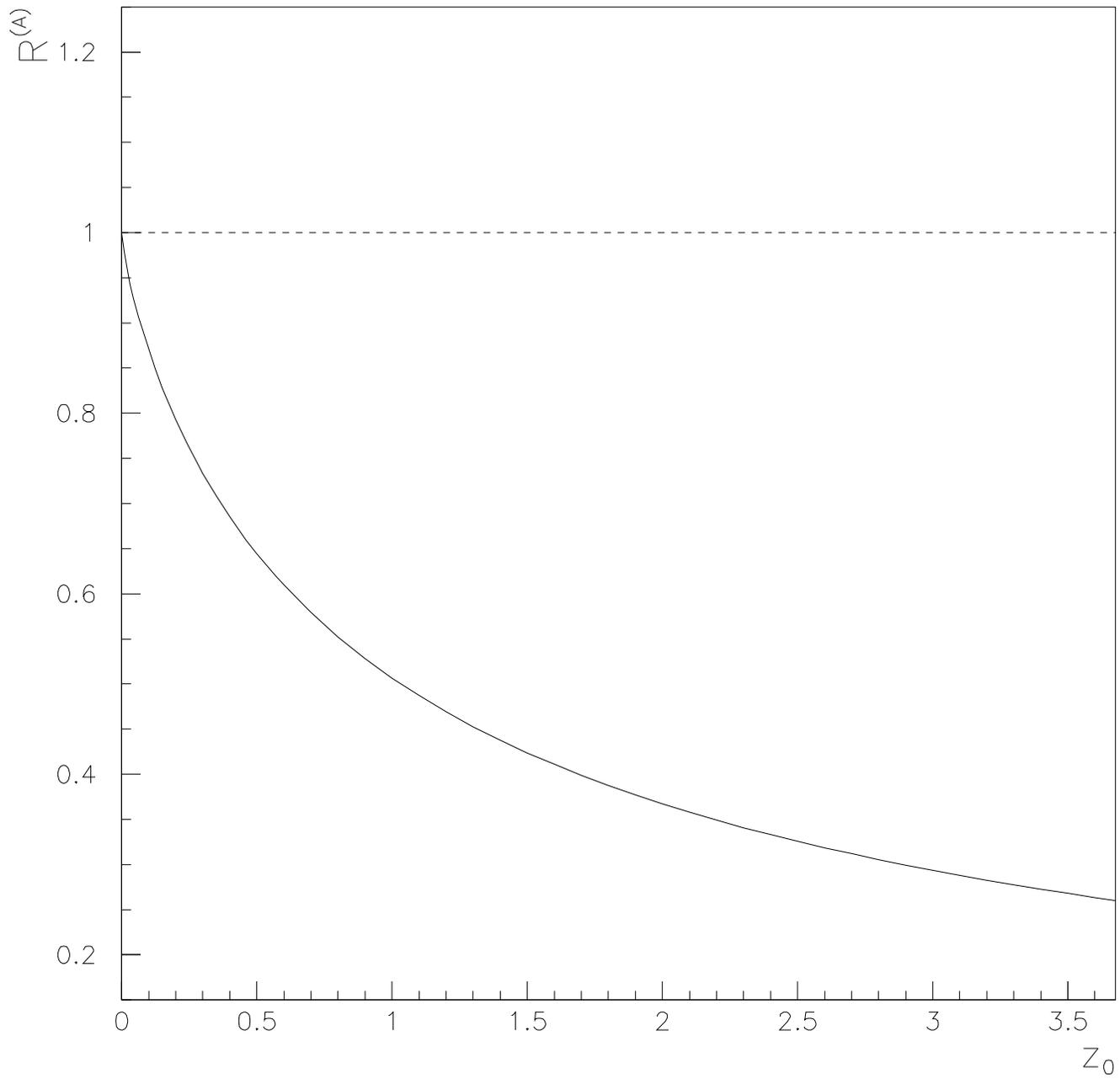
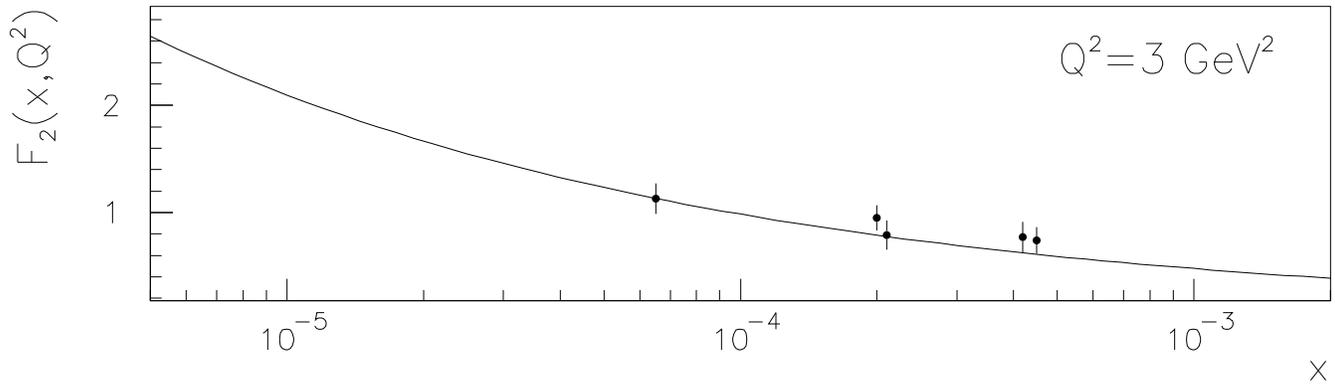
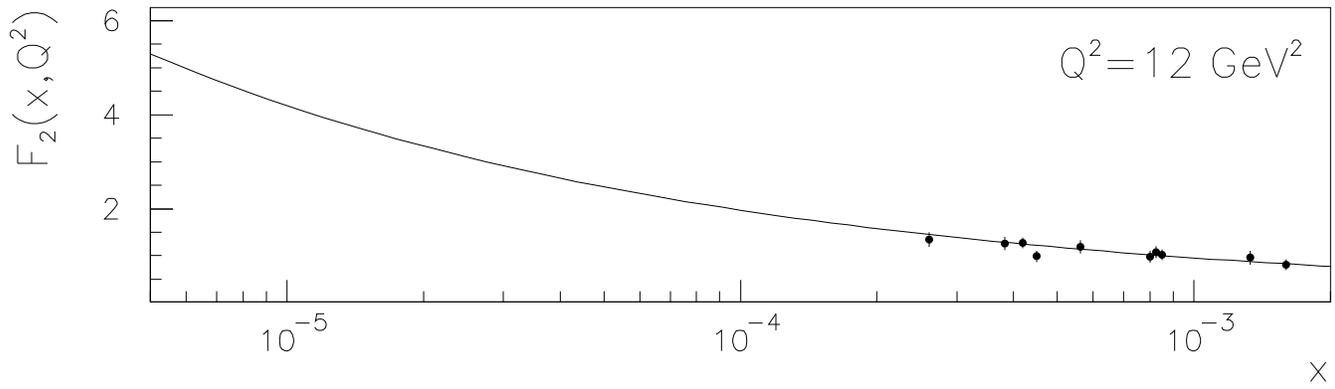
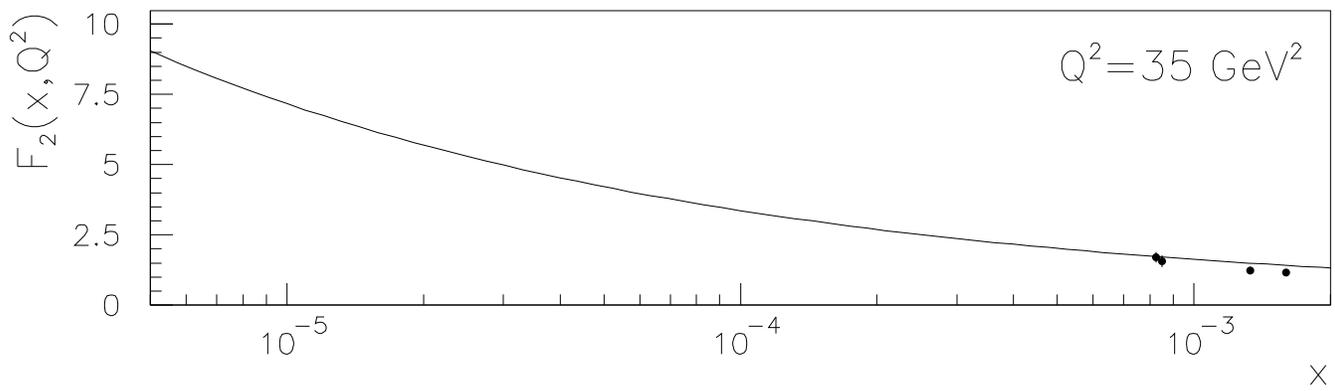
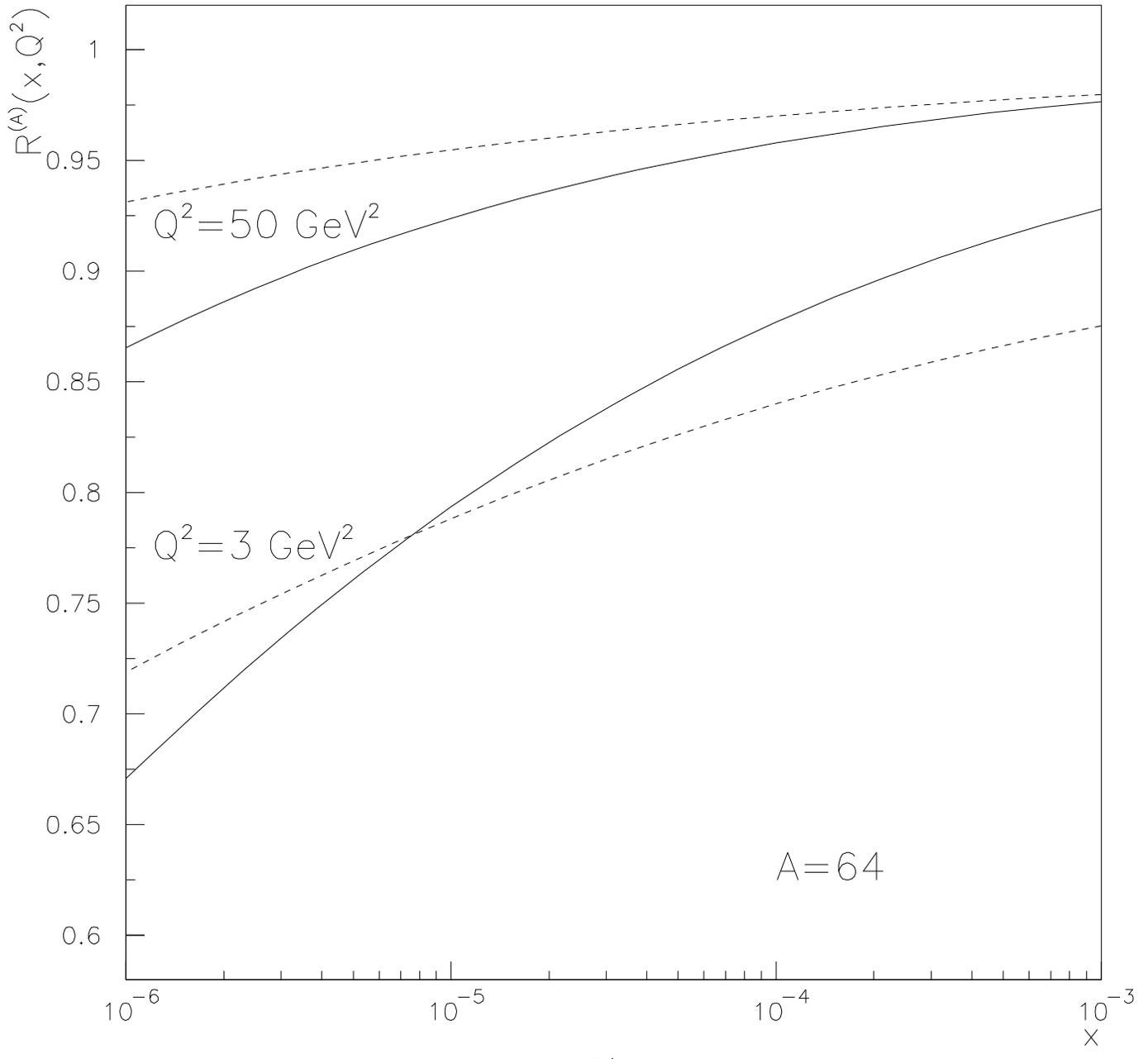
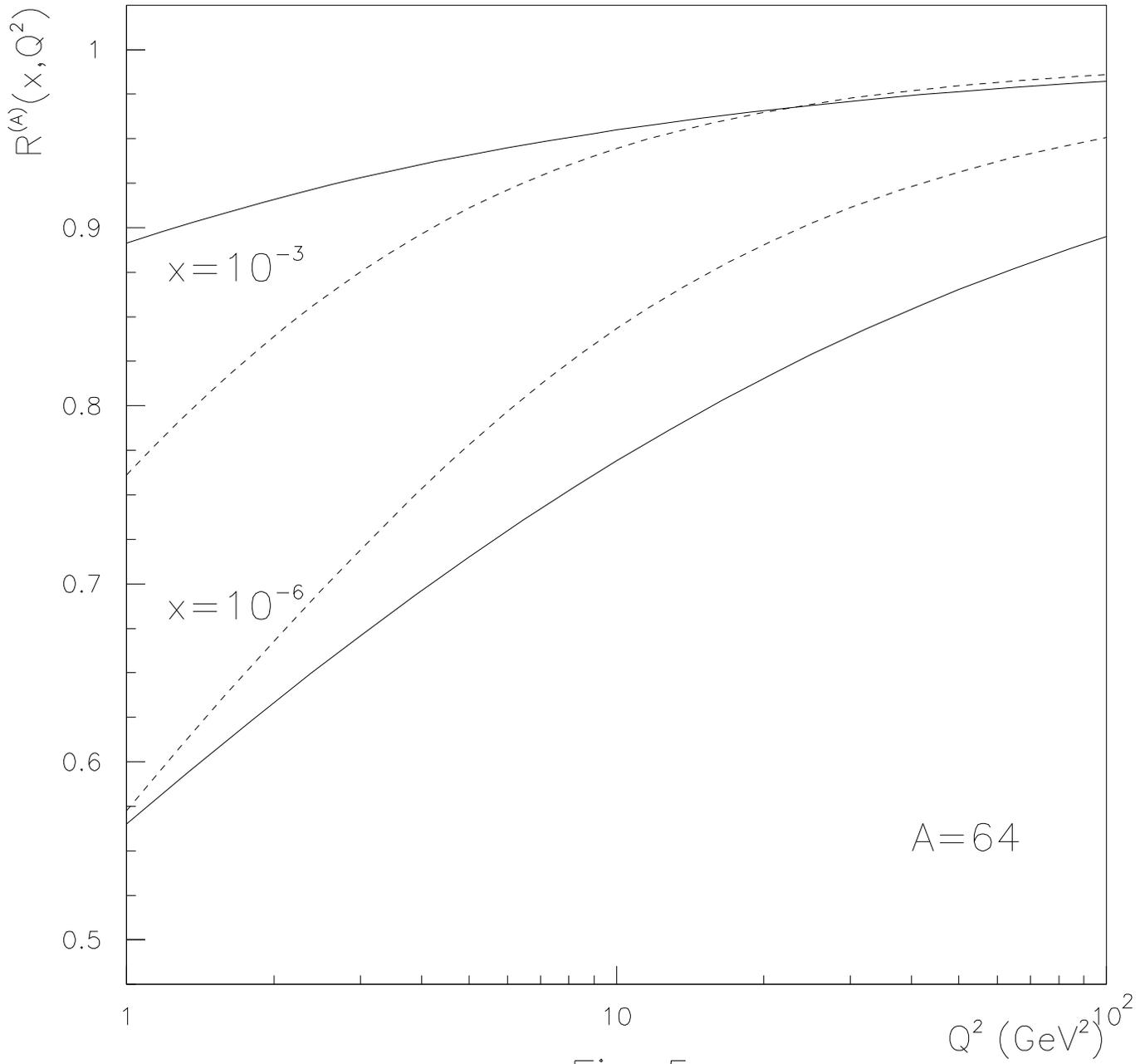
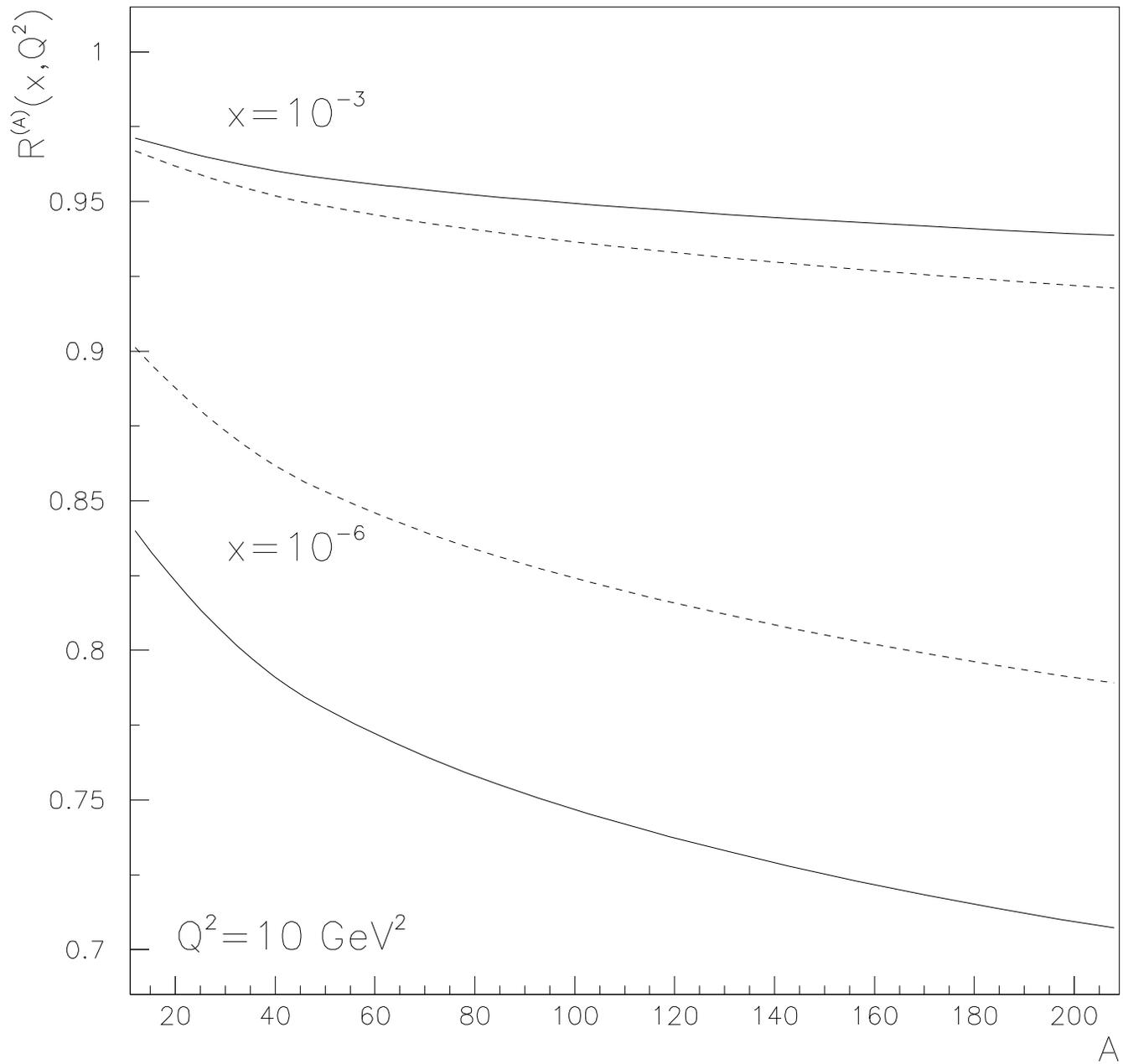
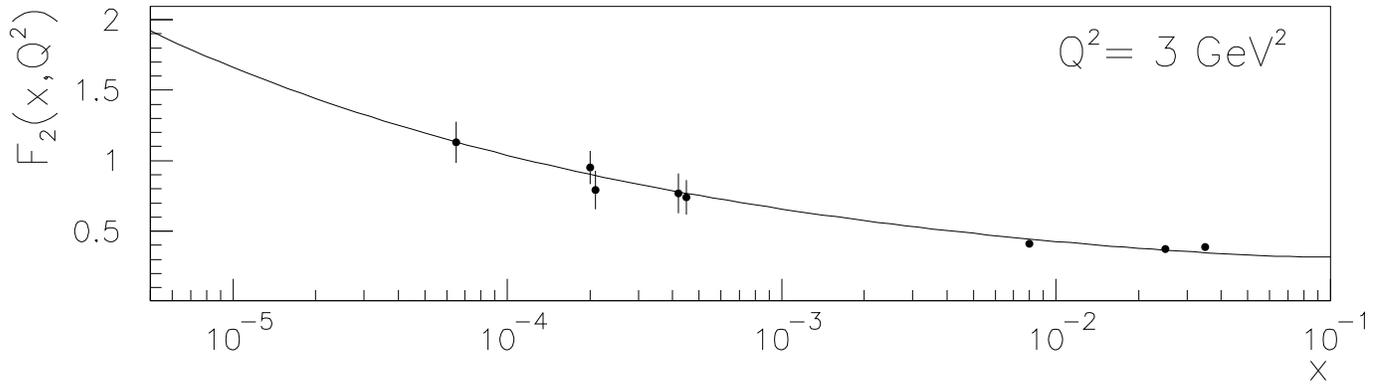
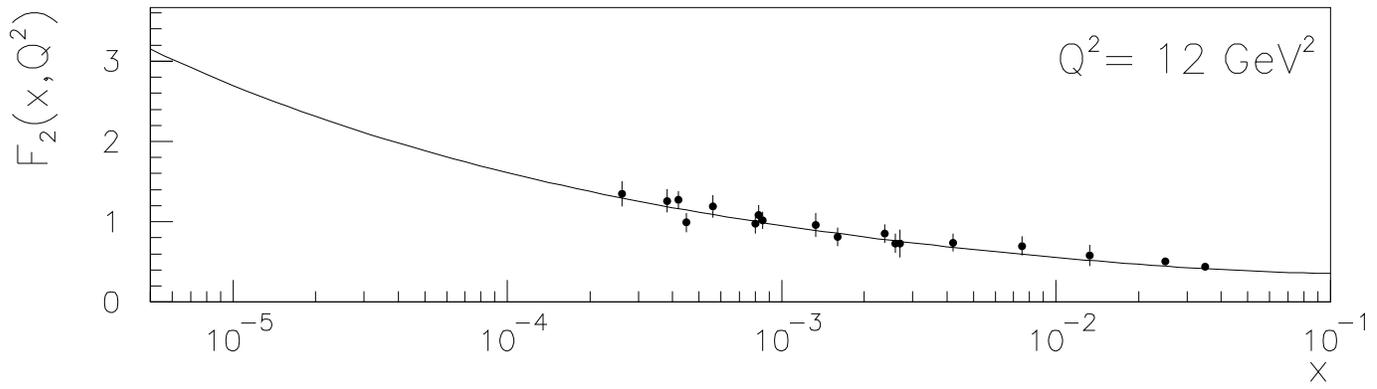
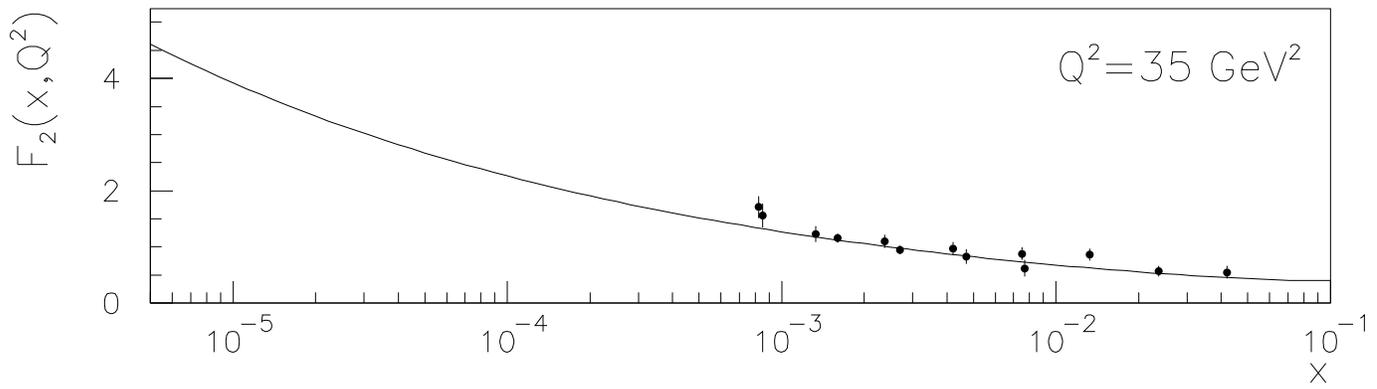
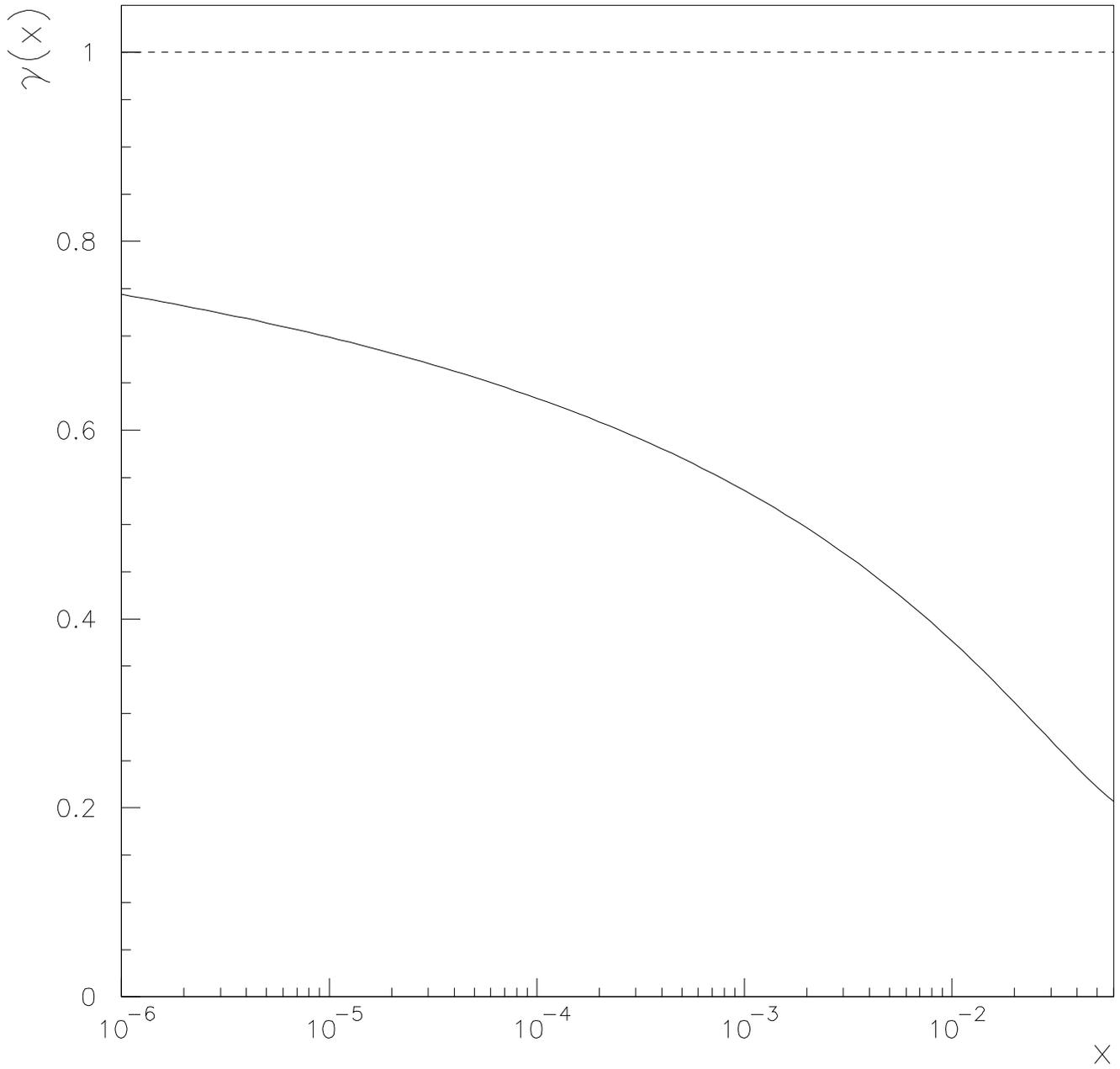
\end{document}